\documentclass[aps,prl,preprint,superscriptaddress]{revtex4}

\usepackage{graphicx}
\usepackage{dcolumn}
\usepackage{bm}
\usepackage{amssymb}
\usepackage{amsmath}

\newcommand{\lco}{LaCoO$_3$}
\newcommand{\lsco}{La$_{0.998}$Sr$_{0.002}$CoO$_3$}

\begin{document}

\title{Spin--state polaron in lightly hole-doped LaCoO$_{3}$}



\author{A.~Podlesnyak}
\thanks{Corresponding author. Electronic address: andrei.podlesnyak@hmi.de}
\author{M.~Russina}
\affiliation{Hahn--Meitner--Institut, Glienicker Stra{\ss}e 100,
Berlin 14109, Germany}
\author{A.~Furrer}
\affiliation{Laboratory for Neutron Scattering, ETH Zurich \& Paul Scherrer
Institut, CH-5232 Villigen PSI, Switzerland}
\author{A.~Alfonsov}
\affiliation{Leibniz Institute for Solid State and Materials Research IFW Dresden, Germany}
\author{E.~Vavilova}
\affiliation{Leibniz Institute for Solid State and Materials Research IFW Dresden, Germany}
\affiliation{Zavoisky Physical Technical Institute of
the Russian Academy of Sciences, 420029 Kazan, Russia}
\author{V.~Kataev}
\author{B.~B\"uchner}
\affiliation{IFW Dresden, P.O. Box 270116, D-01171 Dresden, Germany}
\author{Th.~Str{\"a}ssle}
\affiliation{Laboratory for Neutron Scattering, ETH Zurich \& Paul
Scherrer Institut, CH-5232 Villigen PSI, Switzerland}
\author{E.~Pomjakushina}
\affiliation{Laboratory for Neutron Scattering, ETH Zurich \& Paul
Scherrer Institut, CH-5232 Villigen PSI, Switzerland}
\affiliation{Laboratory for Developments and Methods, Paul Scherrer
Institut, CH-5232 Villigen PSI, Switzerland}
\author{K.~Conder}
\affiliation{Laboratory for Developments and Methods, Paul Scherrer
Institut, CH-5232 Villigen PSI, Switzerland}
\author{D.~I.~Khomskii}
\affiliation{II~Physikalisches Institut, Universit{\"a}t zu
K{\"o}ln, Z{\"u}lpicher Stra{\ss}e 77, 50937 K{\"o}ln, Germany}

\date{\today}

\begin{abstract}
Inelastic neutron scattering (INS), electron spin (ESR) and nuclear
magnetic resonance (NMR) measurements were employed to establish the
origin of the strong magnetic signal in lightly hole-doped
La$_{1-x}$Sr$_{x}$CoO$_{3}$, $x\sim 0.002$. Both, INS and ESR low
temperature spectra show intense excitations with large effective
$g$-factors $\sim 10-18$. NMR data indicate the creation of extended
magnetic clusters. From the $Q$-dependence of the INS magnetic
intensity we conclude that the observed anomalies are caused by the
formation of octahedrally shaped spin-state polarons comprising
seven Co ions.

\end{abstract}


\maketitle

Physical properties of nanostructured magnetic materials are
extensively studied because of their fundamental interest and
potential applications. A naturally occurring analog to the
artificially fabricated heterostructures are doped perovskites with
intrinsic inhomogeneities (magnetic phase separation), i.e. with a
spatial coexistence of magnetic clusters in a nonmagnetic matrix.
Hole-doped cobaltites La$_{1-x}$Sr$_{x}$CoO$_{3}$ with
perovskite-type structure exhibit ferromagnetic (FM) clusters
\cite{Kuhns,Louca2,Phelan} causing spin-glass and superparamagnetic
behavior for $0.05 \lesssim x \lesssim 0.2$ below and above a
critical magnetic blocking temperature $T_g$, respectively
\cite{Senaris}. Due to a progressive change with increasing
temperature from low- (LS) to intermediate- (IS) or high-spin (HS)
states of the cobalt ions a reentrant metal-insulator (MI)
transition was found for $0.2 \lesssim x \lesssim 0.3$ within $100
\lesssim T \lesssim 200$~\,K \cite{Senaris}. With the addition of
charge carriers the number and possibly size of clusters grow
leading to a percolation-type long-range FM order and MI transition
at $x \gtrsim 0.2$ \cite{Phelan,Senaris,Wu1}.

Most of the investigations up to now have been focused on relatively
high Sr concentration ($x>0.1$). It is widely believed that the
addition of each hole into pristine LaCoO$_3$ through the
substitution of a divalent ion for La$^{3+}$ creates a Co$^{4+}$ ion
in the lattice which has a nonzero $S$ in any spin state
configuration, thereby inducing a magnetic moment in the system. An
amazing fact was found by Yamaguchi et al. in 1996 \cite{Yam1} and
apparently forgotten later. Namely, already lightly doped material
with $x\sim0.002$ (i.e. with an estimated concentration of only two
holes per thousand Co$^{3+}$ ions) exhibits unusual paramagnetic
properties at low temperatures: few embedded spins in a nonmagnetic
matrix give an order of magnitude larger magnetic susceptibility
than expected. It was proposed that a doped hole in the spin-singlet
ground state of LaCoO$_3$ behaves as a localized magnetic impurity
with unusually large spin value $S=10-16$ \cite{Yam1} due to the
formation of a magnetic polaron whose nature, however, remained
unclear. Later, and for higher Sr-doping $x>0.05$, it was surmised
that the addition of charge carriers forms Zener-type polarons or
even many-site magnetopolarons \cite{Louca2,Phelan}. However,
experimental proof of the existence of such polarons is missing so
far.

In this Letter, we elucidate the mechanism of how already the light
hole doping $x \sim 0.002$  dramatically affects magnetic properties
of LaCoO$_3$. Combining INS data, obtained with and without external
magnetic field, with the single crystal ESR and NMR measurements on
La$_{0.998}$Sr$_{0.002}$CoO$_3$, we find that the charges introduced
by substitution of Sr$^{2+}$ for La$^{3+}$ do not remain localized
at the Co$^{4+}$ sites. Instead, each hole is extended over the
neighboring Co$^{3+}$ ions, transforming them to higher spin state
and thereby forming a magnetic seven-site (heptamer) polaron.

Highly stoichiometric powder and single crystal samples of
La$_{1-x}$Sr$_x$CoO$_3$, $x=$\,0, 0.002 were synthesized and
characterized according to procedures described elsewhere
\cite{JMMM}. The INS measurements were performed on the
high-resolution time-of-flight spectrometers NEAT
(Hahn-Meitner-Institut, Berlin, Germany) and FOCUS (Paul Scherrer
Institut, Villigen, Switzerland). The data were collected using
incoming neutron energies $3.26-3.5$\,meV, giving an energy
resolution at the elastic position of $\sim 0.09-0.15$\,meV. Raw
data were corrected for sample self-shielding and detector
efficiency according to standard procedures. The DAVE software
package was used for elements of the data reduction and analysis
\cite{Dave}. High field ESR measurements were performed with a
home-made spectrometer based on a Millimeterwave Vector Network
Analyzer (MVNA) from AB Millimetr\`e at frequencies 27\,-\,550\,GHz
for the magnetic field $B$ parallel to the [001] pseudo-cubic  axis
of the single crystal (see technical details in
Ref.~\onlinecite{Golze06}). In the same field geometry $^{59}$Co ($I
=7/2$) NMR was measured at a frequency of 47.65\,MHz with a Tecmag
pulse NMR spectrometer.

\begin{figure}[tb!]
\begin{center}
\includegraphics[width=0.8\columnwidth]{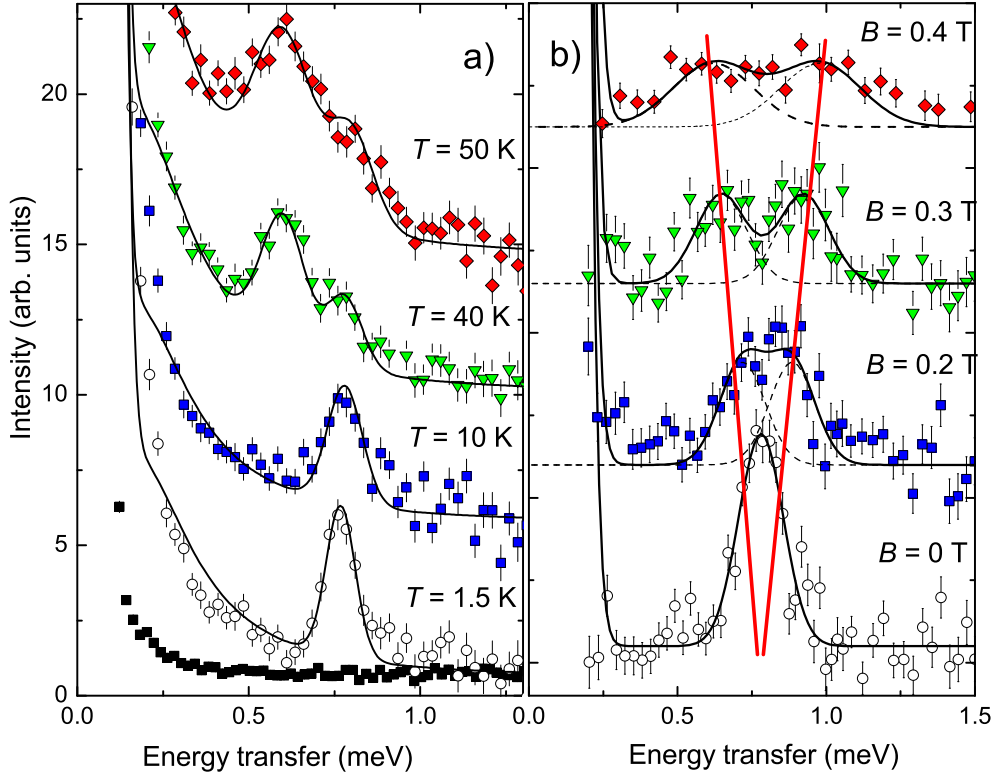}
\caption {(Color online) Temperature ($B=0$) and magnetic field
($T=1.5$\,K) evolutions of the INS spectra of
La$_{0.998}$Sr$_{0.002}$CoO$_3$ measured on FOCUS and NEAT,
respectively. Solid black squares correspond to data taken from
nonmagnetic LaAlO$_{3}$ at $T=50$\,K, black lines refer to
least-squares fits of Gaussian functions, red lines are guides to
the eye.\label{Fig:2}}
\end{center}
\end{figure}

The susceptibility data (not shown) are similar to those of
Ref.~\cite{Yam1}. In order to estimate an effective magnetic moment
of doped holes we fitted measured magnetization $M(H)$ with a
combination of the conventional Brillouin function B$_S(y)$ and a
field-linear term, $M(H) = N \mu_{\textsc{b}} g S \cdot B_S(y) +
\chi_0 H,$~ $y = (g \mu_{\textsc{b}} S H) / (k_{\textsc{b}} T)$.
Assuming a hole concentration $N=0.002$, we found $gS \sim
15\mu_{\textsc{b}}$/hole, which is much larger than we can expect
from Co$^{3+}$ or Co$^{4+}$ in any spin-state, and which agrees with
\cite{Yam1}.

Zero-field inelastic neutron spectra of
La$_{0.998}$Sr$_{0.002}$CoO$_3$ are shown in Fig.~\ref{Fig:2}\,a. In
contrast to the parent compound LaCoO$_{3}$, where no excitations
have been found for $T<30$\,K \cite{Podlesnyak1}, an inelastic peak
at 0.75\,meV was observed down to $T=1.5$\,K. One more inelastic
peak at 0.6\,meV was found at intermediate temperatures starting
from $T \sim 30$\,K similar to that found in pristine LaCoO$_{3}$
\cite{Podlesnyak1}. Clearly the peak at 0.6\,meV corresponds to the
signal from the undisturbed matrix. We can thus interpret the peak
observed at 0.75\,meV as a signal which is due to Sr doping
\cite{JMMM}. Already a weak magnetic field splits the transition
into two lines whose widths widen considerably with increasing field
strength (Fig.~\ref{Fig:2}\,b). The Zeeman splitting is enormous and
can be explained with a $g$-factor of the order of 10 in agreement
with the aforementioned macroscopic measurements.

\begin{figure}[tb!]
\begin{center}
\includegraphics[width=0.7\columnwidth]{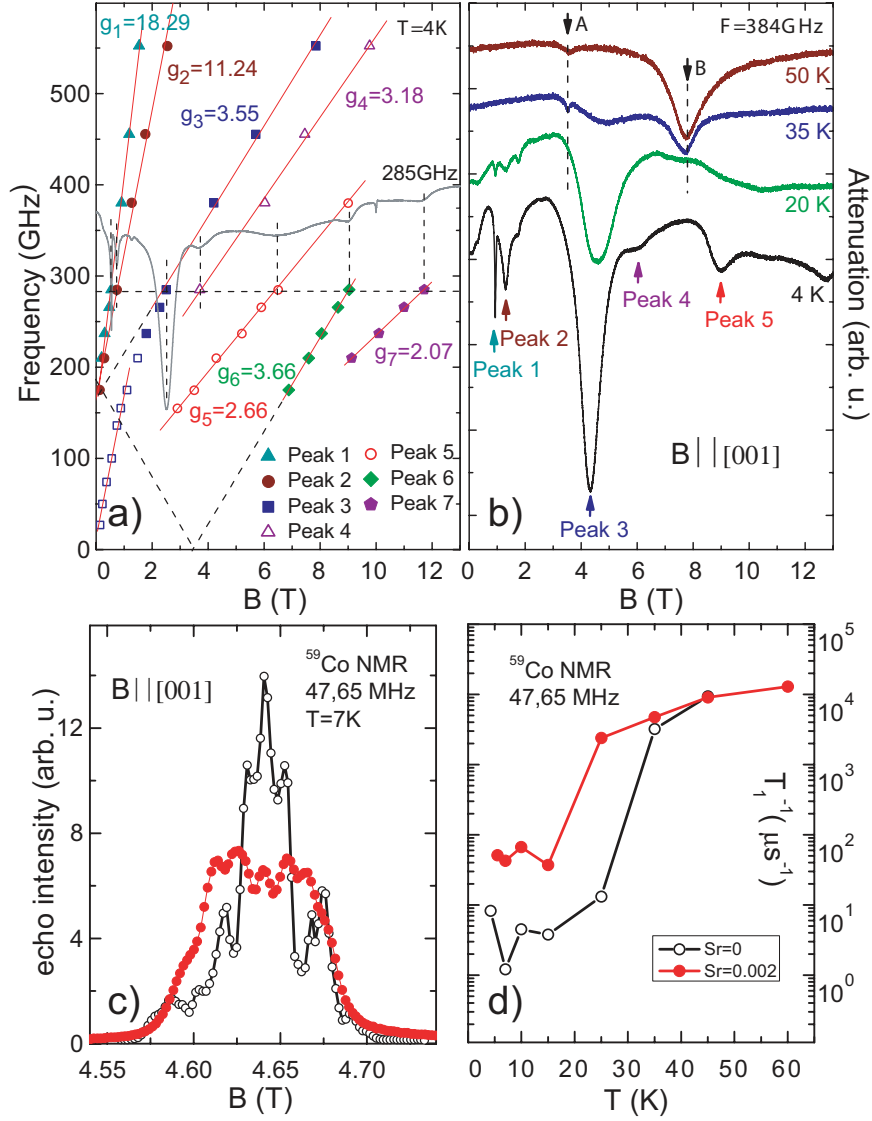}
\caption{(Color online) a) Frequency vs. magnetic field dependence
(branches) of the ESR modes of the low-$T$ spectrum. Straight lines
through data points are linear fits (see text). Open squares denote
a small presumably impurity peak visible below $\sim 200$\,GHz. b)
$T$-dependence of the ESR spectrum at 384\,GHz. A and B label ESR
modes due to thermally activated Co$^{3+}$ HS state ions. c) and d)
Low-$T$ $^{59}$Co NMR spectra and $T$-dependences of the nuclear
relaxation rate $T_1^{-1}$, respectively, for \lco\ and \lsco\
single crystals (open and closed circles). Lines connecting the data
points are guides for the eye.} \label{ESR_NMR}
\end{center}
\end{figure}

Similarly to INS, the undoped \lco\ exhibits no bulk ESR signal for
$T \leq 30$\,K \cite{Noguchi}. However,  in \lsco\ we observe a very
intense ESR spectrum consisting of 7 absorption lines
(Fig.~\ref{ESR_NMR}\,a,b). The dependence of their resonance fields
$B_{res}^{i}$ on the frequency $\nu$ (resonance branches) reveals
that most of the excitations are gapped with a gap value $f_0
\approx 170$\,GHz $\approx 0.7$\,meV (Fig.~\ref{ESR_NMR}\,a,b), in
nice agreement with the energy of the low-$T$ INS peak. The
effective $g$-factors of the most intense branches $g_i = (h/\mu_B)
(\partial\nu/\partial B)_i$ are significantly larger than a
spin-only value of 2 and vary from $\sim 2.1$ to $\sim 18.3$. With
increasing $T$ the intensity of these lines strongly decreases
whereas above $\sim 35$\,K two new lines (marked A and B in
Fig.~\ref{ESR_NMR}\,a,b) emerge. Their branches (not shown) yield a
gap $f_1 \approx 150$\,GHz $\approx 0.6$\,meV and a $g$-factor
$\approx 3.43$. The behavior of A and B is very similar to the ESR
data in Ref.~\onlinecite{Noguchi} for {\it undoped} \lco\ which
allows to unambiguously identify these lines with the thermally
activated Co$^{3+}$ HS state ions and thus with the thermally
activated INS peak \cite{Podlesnyak1}.

The strong low-$T$ ESR response of \lsco\ cannot be explained by the
occurrence of isolated Co$^{4+}$ ions with an effective spin
$\tilde{S}=1/2$ in the LS state or isolated Co$^{3+}$ in the HS or
IS state with $\tilde{S}=1$. A large number of lines implies the
existence of resonating centers with larger spin multiplicity,
since, e.g., for $\tilde{S}=1$ not more than 3 lines can be
expected. Therefore the ESR data strongly suggest that small Sr
(i.e. hole) doping results in the formation of spin clusters with
large effective $g$-factors involving several interacting magnetic
Co sites.

The $^{59}$Co NMR data are summarized in Fig.~\ref{ESR_NMR}\,c,d.
The spectral shape and the spin-lattice relaxation rates of the
undoped \lco\ agree very well with previous $^{59}$Co-NMR studies
\cite{Itoh95,Kobayashi}. According to a simple estimate, doping with
0.2\,\% Sr, that yields 0.2\,\% of Co$^{4+}$ sites, should change
the electric field gradient for at most 5\,\% of nuclei. It means
that the doping induced change of the low-$T$ spectrum, that gets
barely resolved (Fig.~\ref{ESR_NMR}\,c), is not due to the
quadrupole interaction and has probably magnetic origin. It becomes
even more apparent in the nuclear spin dynamics yielding at low $T$
a more than 15 times enhanced relaxation rate $T_1^{-1}$
(Fig.~\ref{ESR_NMR}\,d). The observed stretch-exponential shape of
the nuclear magnetization recovery suggests a substantially
non-uniform distribution of local magnetic environments at low $T$
seen by the Co nuclei \cite{Hoch04}. Thus $^{59}$Co-NMR data of
\lsco\ clearly indicate the formation of spatially extended magnetic
clusters at low $T$. In contrast, above $\sim 35$\,K, where a
considerable part of Co ions is in the thermally activated HS state,
the NMR spectra and relaxation for doped and undoped samples are
very similar, and the shape of the nuclear magnetization recovery
testifies an almost homogeneous distribution of magnetic centers
\cite{NMRcomment}.

Thus, a contribution of several Co-ions, i.e. the formation of
magnetic clusters is required to explain the results of our magnetic
susceptibility, INS, ESR and NMR measurements. The hole is not
localized on one particular ion but dynamically distributed over the
cluster. A reasonable mechanism for such a resonant state was
proposed by Louca and Sarrao \cite{Louca2}. Neighboring LS-Co$^{4+}$
and IS-Co$^{3+}$ ions can share an $e_g$ electron by swapping
configuration. The $t_{2g}$ electrons, in turn, couple
ferromagnetically via double exchange interaction. Therefore, we
propose that the holes introduced in the LS state of LaCoO$_3$ are
extended over the neighboring Co sites forming spin-state polarons
and transforming all involved Co$^{3+}$ ions to the IS state. An
important question remains: How many Co ions are involved in a
single hole-doped cluster?

\begin{figure}[tb!]
\begin{center}
\includegraphics[width=0.5 \columnwidth]{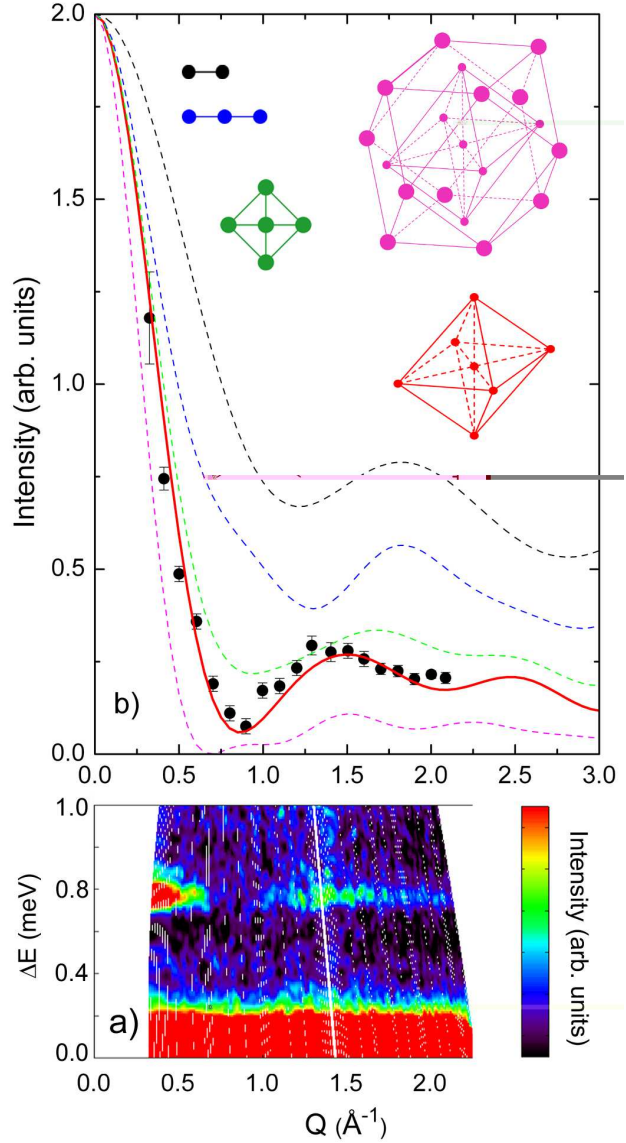}
\caption {(Color) a) Excitation INS spectrum collected on FOCUS from
La$_{0.998}$Sr$_{0.002}$CoO$_3$ at $T=1.5$\,K. b) Circles:
Experimental $Q$ dependence of the intensity of the peak observed at
0.75\,meV. Lines: Calculated $Q$ dependence of the neutron cross
section [Eq.~\ref{average}] for different Co multimers (visualized
in the figure) in the cubic perovskite lattice of LaCoO$_3$ and for
$| S \rangle \Rightarrow | S \rangle$ transitions. The nearest
neighbor Co--Co distance was fixed at
R$_{\textrm{Co-Co}}=3.9$\,\AA~determined for
La$_{1-x}$Sr$_{x}$CoO$_3$ from the Co-O pair density function
\cite{Louca2}. \label{INS}}
\end{center}
\end{figure}

The wave-vector dependence of the intensity of the INS signal yields
direct information about the geometrical configuration of the
magnetic ions in the cluster. We studied in detail the $Q$
dependence of the 0.75\,meV peak for $0.4 \leqslant Q \leqslant
2.0$\,\AA. The excitation is dispersionless indicating that
intercluster interactions can be neglected (Fig.~\ref{INS}\,a). The
intensity of the observed transition exhibits a clear oscillatory
behavior reflecting the size as well as the shape of a spin-state
polaron through the structure factor. For a cluster comprising $n$
magnetic ions, the neutron cross section for polycrystalline
materials is given by \cite{Furrer}:
\begin{equation}
\frac{d^2\sigma}{d\Omega d\omega} \quad\propto\quad F^2(Q)
\sum_{j<j{^\prime}=1}^n \Bigg( | \langle S \| T_j \| S^\prime\rangle
|^2
+ 2\frac{\sin (Q | R_j - R_{j\prime} | )}{Q | R_j - R_{j\prime} |}
\langle S \| T_j \| S^\prime\rangle \langle S^\prime \| T_{j^\prime}
\| S \rangle \Bigg), \label{average}
\end{equation}
where $F(Q)$ is the magnetic form factor, $Q$ the scattering vector,
$R_j$ the position vector of the $j$-th ion in the cluster, and
$T_j$ an irreducible tensor operator of rank 1 \cite{Judd}. This
cross section corresponds to a superposition of damped sine
functions which reflect the geometry of the cluster. Each particular
transition $| S \rangle \Rightarrow | S^\prime \rangle$ has its
specific $Q$ dependence due to both the sign and the size of the
reduced matrix elements. Lines in Fig.~\ref{INS}\,b correspond to
calculated cross sections for different Co clusters in the cubic
approximation of the perovskite lattice of LaCoO$_3$ and for the
special case of a $| S \rangle \Rightarrow | S \rangle$ transition
\cite{noteA}, which, as will be seen below, is relevant in the
context of the present work. We clearly see that the $Q$ dependence
of the cross section is an unambiguous fingerprint of the geometry
of the multimers; in particular, the data observed for the 0.75\,meV
transition in La$_{0.998}$Sr$_{0.002}$CoO$_3$ are perfectly
explained by the scattering from \emph{an octahedrally shaped Co
heptamer} (see Fig.~\ref{INS}\,b, red heptamer and red solid line).
Total moment of this heptamer (consisting formally of one LS
Co$^{4+}$ ($S=1/2$) and six IS Co$^{3+}$ ($S=1$) is 15
$\mu_{\textsc{b}}$, in exact agreement with our magnetic
measurements.

Considering only nearest-neighbor coupling $J$ between a central
Co$^{4+}$ ion in the LS state and six Co$^{3+}$ ions in the IS
state, the Heisenberg exchange Hamiltonian is given by $H_{ex} =
-2J\vec{S}_1 \cdot \vec{S}_A$, where $\vec{S}_A=\vec{S}_2 + \ldots
+\vec{S}_7$ and the total spin is $\vec{S}=\vec{S}_1+\vec{S}_A$. The
Co-Co coupling $J$ is ferromagnetic via the double exchange
mechanism \cite{Louca2}. The ground state of the cobalt heptamer is
therefore the state with maximum spin quantum numbers, namely $\mid
S_1,S_A,S\rangle = \mid 1/2,6,13/2 \rangle$. The first excited state
$\mid 1/2,5,11/2 \rangle$ lies higher up by $J$. The exchange
coupling $J$ of cobalt oxides is of the order of 10\,meV
\cite{Wagner}, thus, the first excited multiplet heptamer state lies
far above the energy window covered by the present experiments. What
is then the origin of the peak observed at 0.75\,meV? As argued
above, even light doping promotes neighboring Co$^{3+}$ to an IS
state already at T=0. The ground state of the Co$^{3+}$ ions in the
IS state comprising the main part of the magnetic polarons is an
orbitally degenerate triplet (see Ref.~\onlinecite{Podlesnyak1},
Fig.~4\,(right)) which is split by a small trigonal field into a
singlet and doublet. Transition between these levels is, in our
opinion, the source of the 0.75~meV peak. In fact, the temperature
dependence of the intensity supports the singlet-doublet nature of
the peak at 0.75\,meV, see Fig.~\ref{Fig:2}\,a. Here, this
transition due to a magnetic (IS) Co$^{3+}$ exists already at $T=0$,
whereas for the undoped system the similar transition at 0.6~meV is
only due to thermally-activated magnetic (HS) state
\cite{Podlesnyak1}. The complete ground-state wave function of the
Co$^{3+}$ heptamer has to be written as a combined spin-orbit
product state of the form $\mid S_1,S_A,S\rangle \mid L,M_L
\rangle$, thus the intensities of both spin and orbital excitations
are governed by the structure factor of the cobalt heptamer
described by Eq.~\ref{average}.

To summarize, we have investigated lightly doped cobaltite
La$_{0.998}$Sr$_{0.002}$CoO$_{3}$ by means of INS, ESR and NMR
techniques. Our work gives a clear microscopic explanation why hole
doping of as little as 0.2\% may dramatically affect the overall
magnetic properties of the entire system. We have found that holes
introduced in the LS state of LaCoO$_3$ by substitution of Sr$^{2+}$
for La$^{3+}$ transform the six nearest neighboring Co$^{3+}$ ions
to the IS state forming octahedrally shaped spin-state polarons. The
formation of spin-state polarons may be a common mechanism present
in other Co-based compounds. Spin-state polarons behave like
magnetic nanoparticles embedded in an insulating nonmagnetic matrix.
Additional charge carriers increase the number of such spin-state
polarons, which form a percolative network resulting in a metallic
state with long-range FM order at the critical concentration
$x_c=0.18$ \cite{Phelan}.

This work is partly based on experiments performed at the Swiss
spallation neutron source SINQ,  Paul Scherrer Institute, Villigen,
Switzerland. We acknowledge support by the European Commission under
the 6th Framework Programme through the Key Action 'Strengthening
the European Research Area, Research Infrastructures' (contract:
RII3-CT-2003-505925), by the European project COMEPHS, by the Swiss
National Science Foundation (SCOPES IB7320-110859/1, NCCR MaNEP) and
by the German-Russian cooperation project of the DFG (grant No. 436
RUS 113/936/0-1), by SFB 608 and of the RFBR (grants No.
08-02-91952-NNIO-a \& No. 07-02-01184-a).

\bibliographystyle{apsrev}

\begin{thebibliography}{19}
\bibitem[{NMR()}]{NMRcomment}
\bibinfo{howpublished}{We note that the smallest Sr doping level (and
  correspondingly concentration of Co$^{4+}$ ions) investigated so far by NMR
  was 5\,\% \cite{Kobayashi}, i.e. 25 times larger than in our case. Such
  concentration is enough to enable most of Co nuclei to sense magnetic moments
  of Co$^{4+}$ ions. In this doping regime these moments begin to interact and
  form a spin-glass state at low $T$ \cite{Kobayashi}.}
\bibitem[{not()}]{noteA}
\bibinfo{howpublished}{For $| S \rangle \Rightarrow | S \rangle$ transitions we
  have $\langle S \| T_j \| S^\prime\rangle = \langle S^\prime \| T_{j\prime}
  \| S \rangle$, thus the reduced matrix elements are simply a common prefactor
  of the cross section.}
\end{thebibliography}

\end{document}